\documentclass{aa}
\usepackage[varg]{txfonts}
\usepackage{graphicx}
\usepackage{natbib}
\usepackage{pdfpages}
\usepackage[margin=10pt,font=small,labelfont=bf,
labelsep=period]{caption}
\bibpunct{(}{)}{;}{a}{}{,} % to follow the A&A style

\usepackage[normalem]{ulem}

\usepackage{etoolbox}
\usepackage{pdfpages}
\makeatletter
\newcommand*{\AM@pagecommandstar}{}
\define@key{pdfpages}{pagecommand*}{\def\AM@pagecommandstar{#1}}
\patchcmd{\AM@output}{\begingroup\AM@pagecommand\endgroup}
{\ifthenelse{\boolean{AM@firstpage}}{\begingroup\AM@pagecommandstar\endgroup}{\begingroup\AM@pagecommand\endgroup}}{}{} % Patch to use new option
\patchcmd{\AM@split@optionsii}{\equal{pagecommand}{\AM@temp}\or}
{\equal{pagecommand}{\AM@temp}\or\equal{pagecommand*}{\AM@temp}\or}{}{}
\makeatother

\begin{document}

\title{New reconstruction of event-integrated spectra (spectral fluences) for major solar energetic particle events
\footnote{The reconstructed fluences in tabulated form and the corresponding best-fit parameters are available in electronic form at the CDS via anonymous ftp
 at cdsarc.u-strasbg.fr (130.79.128.5) or via http://cdsarc.u-strasbg.fr/viz-bin/cat/J/A+A/xx/yy.}}
%\titlerunning{Sensitivity estimate of the cosmogenic isotope method}

\author{S. Koldobskiy\inst{1,2,3}
\and O. Raukunen\inst{4}
\and R. Vainio\inst{4}
\and G.A. Kovaltsov\inst{5}
\and I. Usoskin\inst{1,2}
}

\institute{{University of Oulu, Finland}
\and {National Research Nuclear University MEPhI, Moscow, Russia}
\and {St. Petersburg State University, Russia}
\and {University of Turku, Finland}
\and {A.F. Ioffe Physical-Technical Institute of Russian Academy of Sciences, St.Petersburg, Russia}
}

\date{}

\abstract {}
%Aims
{Fluences of solar energetic particles (SEPs) are not easy to evaluate, especially for high-energy events
 (i.e. ground-level enhancements, GLEs).
Earlier estimates of event-integrated SEP fluences for GLEs were based on partly outdated assumptions and data,
 and they required revisions.
Here, we present the results of a full revision of the spectral fluences for most major SEP events (GLEs) for the period from
 1956\,--\,2017 {using updated low-energy flux estimates along with greatly revisited high-energy flux data and  applying the}
 newly invented reconstruction method including an improved neutron-monitor yield function.  }
%Methods
{Low- and high-energy parts of the SEP fluence were estimated using a revised space-borne/ionospheric data
 and ground-based neutron monitors, respectively.
The measured data were fitted by the modified {Band} function spectral shape.
The best-fit parameters and their uncertainties were assessed using a direct Monte Carlo method.
}
%Results
{A full reconstruction of the event-integrated spectral fluences was performed in the energy range above 30 MeV, parametrised and tabulated for easy use along with estimates of the 68\% confidence intervals.}
%Conclusions
{This forms a solid basis for more precise studies of the physics of solar eruptive events and {the transport} of energetic
 particles in the interplanetary medium, as well as the related applications.}

\keywords{Sun:particle emission - Sun:activity - Sun:flares - solar-terrestrial relations}
\titlerunning{GLE spectra}
\maketitle

\section{Introduction}

In addition to Galactic cosmic rays (GCR) continuously bombarding the Earth with slightly variable flux,
 sporadic solar eruptive events, such as flares and/or coronal mass ejections (CMEs)
  may cause dramatic (by many orders of magnitude) enhancements
 of energetic particle fluxes near Earth, which are called solar energetic particle (SEP) events
 \citep{vainio09,desai_LR_16,klein17}.
SEP events occur quite frequently during maxima and early declining phases of solar-activity cycles and
 very seldom during the minimum phase \citep[e.g.][]{bazilevskaya14}.
Sometimes, eruptive events can be sufficiently energetic to  {accelerate particles to relatively high energy} ($>$400 MeV)
 so that they can penetrate the Earth's {magnetosphere and} atmosphere, inducing atmospheric nucleonic cascades, which can be
 measured by ground-based detectors \citep[e.g,][]{shea12}.
Such events are known as ground-level enhancements (GLEs), which are consequently numbered from
 \#1, which took place in February 1942, to the most recent \#72 in September 2017.
The first four GLEs were recorded by ionisation chambers \citep{forbush46} and cannot be
 quantitatively assessed, but those starting from GLE \#5 (23-Feb-1956) were recorded by the network
 of ground-based neutron monitors (NMs) with a possibility of quantifying their intensities and
 spectral parameters \citep[e.g.][]{mishev18}.
All available information of NM data for GLEs \#5\,--\,72 is collected in the
 International GLE Database\footnote{{https://gle.oulu.fi}} \citep[IGLED, see ][]{usoskin_GLE_20}.

Studies of SEP events are important for different reasons.
On one hand, solar eruptive events are well-observed processes of
 energetic-particle acceleration \citep{vainio18}, which can be studied in detail using a multi-messenger approach,
 complementing particle data with observations in different wavelengths \citep[e.g.,][]{Plainaki14,cliver_SEP_16,kocharov17}.
For this purpose, the peak flux intensity and detailed temporal variability of the particle flux
 are important as signatures of the acceleration process in the solar corona and the
 interplanetary medium \citep[e.g.][]{desai_LR_16,kong17}.
Accordingly, numerous studies were focused on peak fluxes of SEPs and {corresponding} acceleration and transport  processes
 \citep[e.g.][]{kouloumvakos15,kocharov17}.
On the other hand, enhanced fluxes of energetic particles affect the radiation environment
 near the Earth \citep[e.g.][]{webber07,mishev15}, making not only the peak fluxes but also the fluence (event-integrated flux) {and its spectral shape of significant importance},
 especially for extreme events \citep[e.g.][]{cliver20}.
 {We emphasise that SEP fluences can not be used for the detailed study of SEP acceleration processes, because
  (i)  the observations at 1 AU are also modified by transport, and (ii) different energies in the fluence spectrum can be
  dominated by different acceleration mechanisms or by the same mechanism operating under different conditions.
It is evident from the proton time-intensity profiles alone that the fluence at MeV and 10-MeV energies is often
 dominated by acceleration at interplanetary shocks \citep[e.g.][]{reames99}.
 \bibpunct[ ]{(}{)}{,}{a}{}{;}
However, the question is more open at 100-MeV and GeV energies peaking much earlier, with possible contributions
 from flares and/or coronal shocks as the main candidates to account for the acceleration \citep[see][and references therein]{cliver_SEP_16}.
 \bibpunct{(}{)}{;}{a}{}{,}
Even if the same CME-driven shock were responsible for the acceleration of 10-MeV and 1-GeV protons, the former
 would typically be accelerated mainly in the solar wind and the latter in the corona, and there is no reason to suggest
  that the spectral form of the fluence would reveal something common about the accelerator properties.
 }

Composing an event-integrated energy {(or rigidity)} spectrum (spectral fluence) of a SEP event is a difficult task.
Direct measurements of the spectrum covering the high-energy ($>$400 MeV) range can be done with modern space-borne
 magnetic spectrometers
 %\textbf{Please ensure and check throughout the paper that all acronyms and abbreviations are written out at first mention, followed by the \uline{abbreviation or acronym in parentheses} (even if you have already introduced them in the Abstract). After that please use only the abbreviation. Instruments, surveys, or facilities do not need an introduction in the Abstract. Please introduce these in the body of the paper unless they are known only by their acronym. See Sect. 5.2.4 of the Language Guide (https://www.aanda.org/for-authors/language-editing/1-introduction) }
 PAMELA
 \citep[Payload for Antimatter Matter Exploration and Light-nuclei Astrophysics --][]{adriani14}
 operated from 2006\,--\,2016, and AMS-02 \citep[Alpha Magnetic Spectrometer --][]{aguilar_AMS_18},
  which has been in operation since 2011.
SEP events analysed using PAMELA data were presented by \citet{bruno18}, while data from AMS-02
 are not available yet \citep{bindi17}.
However, these instruments are located aboard low-orbiting satellites, and thus spend most of their time
 inside the Earth's magnetosphere {and {can  detect  low-energy solar particles only intermittently} (5\,--\,10 minutes per half-orbit),
 leading to essential uncertainties in both SEP peak fluxes and fluences, especially for impulsive events.}
Additionally, uncertainties of the SEP fluxes were large during the earlier years
 \citep[e.g.][]{reeves92,tylka97} due to the saturation of the detectors by strong particle fluxes and the possible penetration of high-energy particles into the detector through the walls of the collimator,
 leading to an enhanced effective acceptance.

Thus, for most of the events, one has to combine data from different instruments, including low-energy
 ($<$400 MeV) space-borne detectors located beyond the magnetosphere, and energy-integrating (above 400 MeV) ground-based NMs.
The reconstruction of the spectrum is typically done by fitting a prescribed spectral shape to the
 data and finding the best-fit parameters of this shape along with their uncertainties.
The first consistent reconstruction of combined spectral fluences for major GLE events was made by
 \citet{tylka09} and updated by \citet{raukunen18}.
It was based on a combination of lower-energy data from different space-borne detectors and the
 high-energy tail based on ground-based NM datasets.
The spectral fluence was estimated by fitting the{ Band-function spectral shape} \citep[double power law with an exponential junction --][]{band93} to the integral rigidity
 spectral fluence.
The space-borne data were collected from different sources, while NM data were analysed by applying
 the NM yield-function by \citet{clem00}.
Spectral fluences were presented as tabulated parameters of the Band-function approximation
 for each GLE event.
This dataset has been extensively used in numerous studies for solar and space physics {\citep{cliver20,Anastasiadis2019,Herbst2019}}, but it has become
 obsolete and requires an essential revision.
First, the high-energy space-borne data have been essentially revisited
%\textbf{is this what you did? - '...it was essential for us to revisit and correct the high-energy space-borne data for known errors.' The specific steps you took for this particular research should be in the simple past: We used, We extrapolated, They determined, This information was analysed etc.  If you are speaking about universal truths, constants, or findings, the present simple should be used: We find, The solution to the equation is, etc. If the research is ongoing, then the present perfect can be used: We have used, We have extrapolated, They have found, etc.  See Sect. 4 of the Language Guide (https://www.aanda.org/for-authors/language-editing/1-introduction). Please do check your past/present tenses carefully throughout, as I can't always be certain that I''ve interpreted your meaning accurately, or it may be ambiguous, and so I risk altering your meaning when I do make changes, and leaving a mistake when I don't. Thank you.\\  }
and corrected for
 known errors \citep{raukunen20}.
Second, the data of the NM network for all GLE events have been revisited \citep{usoskin_GLE_20}{:
 apparent errors were corrected, and the variable GCR background was taken into account}. Furthermore,
 a {most recent and directly verified  using AMS-02 data NM yield function}  \citep{mishev13,mishev20} and a new \textit{\emph{effective-energy}} analysis
 method {were} developed \citep{koldobsky_Eff_2018,koldobsky_Reff_19}.
Moreover, the {Band} function is not an optimum parametric shape for
 the GLE energy {(or rigidity)} spectrum, which requires a roll-off at the highest energies.
Here, we present a complete revision of the reconstructions of the SEP spectral fluences for major SEP events (with GLE)
 using most up-to-date knowledge of the SEP measurements on ground and in space, new models, and a modified spectral form including
 a roll-off at high energies.
 %\textbf{Please note that single-sentence paragraphs are not permitted (excluding your abstract \& any section intros).}

\section{Datasets}

All sources of the energy/rigidity integral fluences used in this study are described below.

\subsection{Space-borne and ionospheric data}
\label{Ss:sat}
Data for the rigidities below 1 GV (energy $<$430 MeV) were taken from space-borne or ionospheric (in the pre-satellite era) data.
For the period since 1989 (i.e. GLEs \#~40\,--\,72), we used all publicly available data from the 
%\textbf{see note 1}
GOES (Geostationary Operational
 Environmental Satellite) energetic particle sensor (EPS) and high energy proton and alpha detector (HEPAD) datasets\footnote{https://www.ngdc.noaa.gov/stp/satellite/goes/index.html} {\citep{Onsager1996,Sellers1996}}.
The fluences at the low-energy channels in this study, that is, $>$30 MeV, $>$50 MeV, $>$60 MeV, and $>$100 MeV, were calculated directly
 from the EPS dataset, but the higher energy HEPAD data were revised using a so-called bow-tie method \citep{vanallen74}.
The nominal HEPAD channels are wide in energy, and they have responses that vary significantly within the channels, and in
 some cases even outside the nominal channel range.
In the bow-tie analysis, calibrated channel responses were folded with an assumed spectral form (power-law), and by varying
 the spectral index within a realistic range for SEP events, optimal effective-energy and geometric-factor values were found for each channel.
The method is explained in full detail in \citet{raukunen20}.
In addition, the cleaning of data spikes (single points with spuriously increased flux) and background subtraction was performed on both GOES datasets.
We note that energy bounds of the HEPAD channels changed slightly after 1995 (GLE \# 53) due to differences in the calibration
 procedures used for GOES-6 (until 1995) and GOES-8 onwards (after 1995).
GLEs \# 71 and 72 were measured by GOES-13, with 10\,--\,100 MeV channels represented by two detectors facing east and west, so that
  the fluences were taken as the average of these two measurements.
Resulting omnidirectional fluences (in units of $10^5$~cm$^{-2}$) are presented in Table~\ref{T:GOES_data}.
{Data channels with rigidity above 1 GV were not used in the fitting procedure (Section~\ref{Sec:fit}) because of their lower reliability.}

We also used the SEP spectrum measured by the PAMELA experiment\footnote{{Experimental data is available at the SSDC cosmic ray database:  {https://tools.ssdc.asi.it/CosmicRays/.}}} \citep{bruno18} for the GLE \#71 (17-May-2012).
For years before 1989, we used fluences from several sources based on different spacecraft and experiments
 \citep{king74,reedy77,goswami88,feynman90,jun07,webber07}.
The exact sources of the low-energy data for each event are specified in the legends of the panels of the figure in Appendix~\ref{App:A}.

\begin{table*}
        \centering
        \caption{Event-integrated omnidirectional integral fluences $F(>E)$ (in units of 10$^5$~cm$^{-2}$) obtained here for GLEs \# 40\,--\,72.
        Columns 1\,--\,5 provide the GLE number, source of data, {start date and time of the event, day of the GCR background calculation
        (BG; BG year is the same as for  start date and is not shown)}, and the integration time (IT).
        Columns 6\,--13 provide the $F(>E)$ values.
        Energies $E$ for different channels are given in MeV.
        Uncertainties include only {statistical, background }subtraction, and bow-tie-method errors, while systematic and instrumental errors
        are not considered.
        Errors smaller than 1\% are not shown.
        $F(>10)$ {and} $F(>462/486)$ channels are not used in the fitting procedure.}
\label{T:GOES_data}
\scalebox{0.93}{
        \begin{tabular}{c|c|c|c|c|r|r|r|r|r|r|r|r}
                \hline
        GLE  & Source & Start date&BG &  IT,  & \multicolumn{8}{c}{$F(>E)$, 10$^5$~cm$^{-2}$}  \\
        \cline{6-13}
        \#  & of data &and time UT & & h & 10 & 30 & 50 & 60 & 100 &336  & 395  & 486\\
        \hline
40      &       GOES-6  &       08:00   25/07/1989      &               24/07   &       24      &$      91.79                                   $&$     68.36                                   $&$     53.83                                   $&$     43.95                                   $&$     19.74                                   $&$     1.90    ^{+     0.06    }_{-    0.06    }               $&$     0.99    ^{+     0.06    }_{-    0.09    }               $&$     0.57    ^{+     0.03    }_{-    0.05    }               $ \\
41      &       GOES-6  &       00:00   16/08/1989      &               24/07   &       24      &$      5024                                    $&$     1666                                    $&$     912.9                                   $&$     638.9                                   $&$     227.3                                   $&$     21.24   ^{+     0.39    }_{-    0.37    }               $&$     10.44   ^{+     0.45    }_{-    0.84    }               $&$     5.83    ^{+     0.22    }_{-    0.44    }               $ \\
42      &       GOES-6  &       11:00   29/09/1989      &               28/09   &       22      &$      10387                                   $&$     7159                                    $&$     5330                                    $&$     4158                                    $&$     1672                                    $&$     145.82  ^{+     2.63    }_{-    2.49    }               $&$     69.48   ^{+     2.99    }_{-    5.56    }               $&$     40.51   ^{+     1.46    }_{-    3.00    }               $ \\
43      &       GOES-6  &       12:00   19/10/1989      &               28/09   &       24      &$      6897                                    $&$     4104                                    $&$     3628                                    $&$     3164                                    $&$     1480                                    $&$     113.20  ^{+     2.05    }_{-    1.94    }               $&$     51.33   ^{+     2.21    }_{-    4.11    }               $&$     27.74   ^{+     1.00    }_{-    2.06    }               $ \\
44      &       GOES-6  &       17:00   22/10/1989      &               28/09   &       24      &$      16241                                   $&$     8199                                    $&$     5373                                    $&$     3741                                    $&$     1149                                    $&$     54.84   ^{+     1.00    }_{-    0.94    }               $&$     22.15   ^{+     0.96    }_{-    1.78    }               $&$     11.30   ^{+     0.41    }_{-    0.84    }               $ \\
45      &       GOES-6  &       17:00   24/10/1989      &               28/09   &       24      &$      7938                                    $&$     3675                                    $&$     2721                                    $&$     2074                                    $&$     839.2                                   $&$     101.70  ^{+     1.84    }_{-    1.74    }               $&$     53.11   ^{+     2.29    }_{-    4.25    }               $&$     31.69   ^{+     1.15    }_{-    2.35    }               $ \\
46      &       GOES-6  &       06:00   15/11/1989      &               14/11   &       24      &$      55.44                                   $&$     43.31                                   $&$     39.86                                   $&$     35.47                                   $&$     15.26                                   $&$     1.31    ^{+     0.05    }_{-    0.05    }               $&$     0.67    ^{+     0.04    }_{-    0.06    }               $&$     0.36    ^{+     0.03    }_{-    0.03    }               $ \\
47      &       GOES-6  &       22:00   21/05/1990      &               05/05   &       24      &$      477.7                                   $&$     271.0                                   $&$     213.5                                   $&$     173.8                                   $&$     72.84                                   $&$     6.31    ^{+     0.13    }_{-    0.12    }               $&$     3.00    ^{+     0.14    }_{-    0.24    }               $&$     1.68    ^{+     0.07    }_{-    0.13    }               $ \\
48      &       GOES-6  &       20:00   24/05/1990      &               05/05   &       24      &$      317.8                                   $&$     235.3                                   $&$     211.8                                   $&$     185.0                                   $&$     90.80                                   $&$     12.07   ^{+     0.23    }_{-    0.22    }               $&$     6.13    ^{+     0.27    }_{-    0.49    }               $&$     3.65    ^{+     0.14    }_{-    0.27    }               $ \\
49      &       GOES-6  &       20:00   26/05/1990      &               05/05   &       24      &$      182.3                                   $&$     144.7                                   $&$     131.0                                   $&$     114.5                                   $&$     55.14                                   $&$     6.49    ^{+     0.13    }_{-    0.13    }               $&$     3.25    ^{+     0.15    }_{-    0.26    }               $&$     1.90    ^{+     0.08    }_{-    0.14    }               $ \\
50      &       GOES-6  &       04:00   28/05/1990      &               05/05   &       24      &$      124.5                                   $&$     102.7                                   $&$     90.42                                   $&$     77.23                                   $&$     37.57                                   $&$     4.38    ^{+     0.10    }_{-    0.09    }               $&$     2.15    ^{+     0.10    }_{-    0.18    }               $&$     1.27    ^{+     0.05    }_{-    0.10    }               $ \\
51      &       GOES-6  &       01:00   11/06/1991      &               31/07   &       11      &$      894.3                                   $&$     559.2                                   $&$     397.3                                   $&$     276.7                                   $&$     80.81                                   $&$     4.09    ^{+     0.09    }_{-    0.08    }               $&$     1.79    ^{+     0.08    }_{-    0.15    }               $&$     1.04    ^{+     0.04    }_{-    0.08    }               $ \\
52      &       GOES-6  &       08:00   15/06/1991      &       31/07   &       24      &$      2883                                    $&$     1215                                    $&$     819.2                                   $&$     595.8                                   $&$     207.3                                   $&$     17.16   ^{+     0.32    }_{-    0.30    }               $&$     8.03    ^{+     0.35    }_{-    0.65    }               $&$     4.54    ^{+     0.17    }_{-    0.34    }               $ \\
53      &       GOES-6  &       19:00   25/06/1992      &               24/06   &       24      &$      680.7                                   $&$     216.9                                   $&$     131.4                                   $&$     88.45                                   $&$     27.69                                   $&$     2.31    ^{+     0.07    }_{-    0.07    }               $&$     1.08    ^{+     0.06    }_{-    0.10    }               $&$     0.63    ^{+     0.04    }_{-    0.05    }               $ \\
\hline          \hline
\multicolumn{5}{c}{} & \multicolumn{8}{c}{$F(>E)$,  10$^5$~cm$^{-2}$}   \\
\cline{6-13}
\multicolumn{5}{c}{} & 10 & 30  & 50    &  60   & 100 & 337  & 392  & 462       \\      
\hline
55      &       GOES-8  &       12:00   06/11/1997      &               01/11   &       24      &$      3165                                    $&$     1248                                    $&$     652.5                                   $&$     496.6                                   $&$     226.7                                   $&$     15.63   ^{+     0.13    }_{-    0.12    }               $&$     8.54    ^{+     0.24    }_{-    0.48    }               $&$     5.42    ^{+     0.13    }_{-    0.24    }               $ \\
56      &       GOES-8  &       13:00   02/05/1998      &               15/05   &       24      &$      471.8                                   $&$     162.8                                   $&$     77.84                                   $&$     57.32                                   $&$     24.60                                   $&$     1.47    ^{+     0.06    }_{-    0.05    }               $&$     0.76    ^{+     0.05    }_{-    0.06    }               $&$     0.51    ^{+     0.04    }_{-    0.04    }               $ \\
58      &       GOES-8  &       20:00   24/08/1998      &               16/08   &       18      &$      683.3                                   $&$     111.6                                   $&$     43.01                                   $&$     31.25                                   $&$     13.18                                   $&$     1.20    ^{+     0.05    }_{-    0.04    }               $&$     0.72    ^{+     0.04    }_{-    0.06    }               $&$     0.42    ^{+     0.03    }_{-    0.04    }               $ \\
59      &       GOES-8  &       10:00   14/07/1900      &               10/07   &       18      &$      47291                                   $&$     18726                                   $&$     7519                                    $&$     4657                                    $&$     1561                                    $&$     59.94   ^{+     0.38    }_{-    0.33    }               $&$     30.26   ^{+     0.82    }_{-    1.70    }               $&$     17.65   ^{+     0.40    }_{-    0.76    }               $ \\
60      &       GOES-8  &       13:00   15/04/2001      &               24/03   &       24      &$      4510                                    $&$     1352                                    $&$     728.0                                   $&$     583.3                                   $&$     302.2                                   $&$     46.07   ^{+     0.30    }_{-    0.26    }               $&$     28.30   ^{+     0.77    }_{-    1.59    }               $&$     19.24   ^{+     0.43    }_{-    0.83    }               $ \\
61      &       GOES-8  &       02:00   18/04/2001      &               24/03   &       24      &$      1509                                    $&$     382.5                                   $&$     168.9                                   $&$     125.0                                   $&$     54.70                                   $&$     6.57    ^{+     0.07    }_{-    0.07    }               $&$     4.13    ^{+     0.12    }_{-    0.24    }               $&$     2.82    ^{+     0.07    }_{-    0.13    }               $ \\
62      &       GOES-8  &       16:00   04/11/2001      &               03/11   &       24      &$      23264                                   $&$     6919                                    $&$     2366                                    $&$     1307                                    $&$     339.9                                   $&$     9.47    ^{+     0.09    }_{-    0.08    }               $&$     4.20    ^{+     0.12    }_{-    0.24    }               $&$     2.28    ^{+     0.06    }_{-    0.11    }               $ \\
63      &       GOES-8  &       04:00   26/12/2001      &               25/12   &       24      &$      3129                                    $&$     790.6                                   $&$     326.4                                   $&$     218.1                                   $&$     77.38                                   $&$     3.25    ^{+     0.05    }_{-    0.05    }               $&$     1.65    ^{+     0.06    }_{-    0.10    }               $&$     0.99    ^{+     0.04    }_{-    0.05    }               $ \\
64      &       GOES-8  &       01:00   24/08/2002      &               10/08   &       24      &$      2273                                    $&$     430.8                                   $&$     178.2                                   $&$     126.5                                   $&$     50.59                                   $&$     3.46    ^{+     0.05    }_{-    0.05    }               $&$     1.81    ^{+     0.06    }_{-    0.11    }               $&$     1.17    ^{+     0.04    }_{-    0.06    }               $ \\
65      &       GOES-10 &       11:00   28/10/2003      &               21/10   &       19      &$      55516                                   $&$     19857                                   $&$     7083                                    $&$     3932                                    $&$     979.1                                   $&$     20.12   ^{+     0.15    }_{-    0.13    }               $&$     10.02   ^{+     0.28    }_{-    0.56    }               $&$     5.88    ^{+     0.14    }_{-    0.26    }               $ \\
66      &       GOES-10 &       20:00   29/10/2003      &               21/10   &       14      &$      8614                                    $&$     2847                                    $&$     1265                                    $&$     889.9                                   $&$     342.0                                   $&$     15.76   ^{+     0.12    }_{-    0.11    }               $&$     8.18    ^{+     0.23    }_{-    0.46    }               $&$     4.83    ^{+     0.12    }_{-    0.21    }               $ \\
67      &       GOES-10 &       17:00   02/11/2003      &               21/10   &       24      &$      5729                                    $&$     1367                                    $&$     461.0                                   $&$     280.4                                   $&$     91.91                                   $&$     5.13    ^{+     0.06    }_{-    0.06    }               $&$     2.73    ^{+     0.09    }_{-    0.16    }               $&$     1.63    ^{+     0.05    }_{-    0.08    }               $ \\
69      &       GOES-11 &       06:00   20/01/2005      &               12/01   &       24      &$      6977                                    $&$     3572                                    $&$     2116                                    $&$     1656                                    $&$     785.4                                   $&$     106.83  ^{+     0.67    }_{-    0.56    }               $&$     63.81   ^{+     1.73    }_{-    3.58    }               $&$     39.16   ^{+     0.87    }_{-    1.69    }               $ \\
70      &       GOES-11 &       02:00   13/12/2006      &               04/12   &       24      &$      3550                                    $&$     1553                                    $&$     796.3                                   $&$     569.7                                   $&$     227.1                                   $&$     15.22   ^{+     0.12    }_{-    0.11    }               $&$     9.04    ^{+     0.25    }_{-    0.51    }               $&$     5.64    ^{+     0.14    }_{-    0.25    }               $ \\
71      &       GOES-13 &       01:00   16/05/2012      &               15/05   &       24      &$      685.2                                   $&$     243.7                                   $&$     123.8                                   $&$     92.11                                   $&$     39.26                                   $&$     2.32    ^{+     0.05    }_{-    0.05    }               $&$     1.80    ^{+     0.07    }_{-    0.11    }               $&$     1.22    ^{+     0.05    }_{-    0.07    }               $ \\
72      &       GOES-13 &       16:00   10/09/2017      &               02/09   &       24      &$      9315                                    $&$     3383                                    $&$     1509                                    $&$     1048                                    $&$     395.0                                   $&$     11.45   ^{+     0.10    }_{-    0.09    }               $&$     7.50    ^{+     0.22    }_{-    0.43    }               $&$     4.37    ^{+     0.11    }_{-    0.20    }               $ \\
%       & 30    & 50    &  60   & 100 & 337  & 392  & 462       \\      \\
                \hline
        \end{tabular}
}
\end{table*}

\subsection{Ground-based data}
\label{Ss:NM}
Data for rigidity above 1 GV (energy $>$430 MeV) were used from a recent reconstruction of
 SEP fluences based on the detrended data of the NM network from IGLED \citep{usoskin_GLE_20}.
For some events and some NMs (mostly low-latitude ones), only an upper limit of the enhancement
 can be set (see vertical blue bars in Figure~\ref{Fig:fit}).
A robust estimation of the SEP fluence using NM data requires the accurate identification of the GLE signal itself.
Since GLEs \# 6, 7, 9, 14, 15, 17, 34, 54, 57, and 68 were weak and lacking sufficiently clear signals,
 we did not analyse these events.  {Originally, the SEP event integration period was deduced from high-energy NM de-trended data \citep{usoskin_GLE_20} and was applied to the low-energy data where we had this opportunity (i.e.  starting from GLE \#40). For earlier GLEs, we used all available data, which sometimes have bigger integration periods. It was taken into account as an additional uncertainty in Step 2 below.}

\section{Fitting procedure}
\label{Sec:fit}

\begin{figure}[t!]
\centering
\includegraphics[width=1\columnwidth]{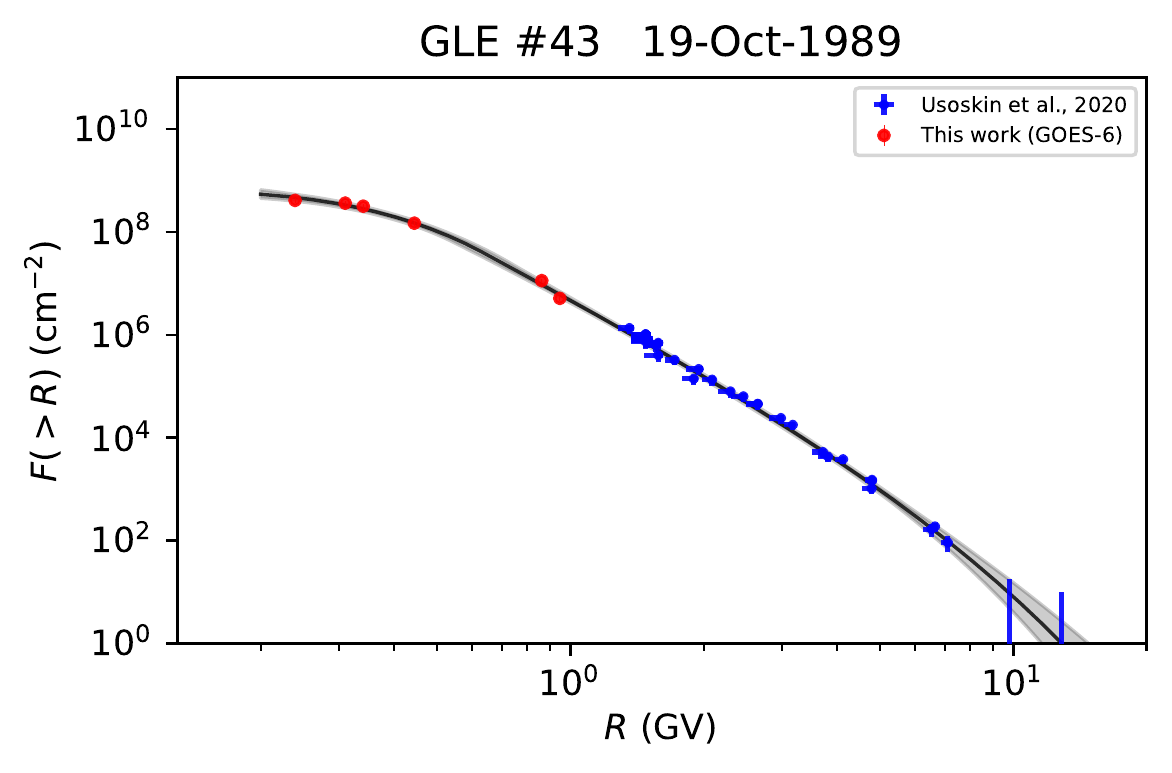}
\caption{Example of the fitting of the modified {Band} function (Equations~\ref{Eq:MBF_1}--\ref{Eq:MBF_2})
 to the data points (Section~\ref{Sec:fit}) for GLE \#43, 19-Oct-1989.
 Blue crosses and vertical lines denote the high-energy NM data with uncertainties, and the upper limits, respectively,
  \citep{usoskin_GLE_20}, while red circles correspond to low-energy data {from GOES-6} (see Table~\ref{T:GOES_data}).
  The grey line with shading depicts the best-fit MBF, along with its 68\% confidence interval.
  }
\label{Fig:fit}
\end{figure}

For each event, the integral fluence points in both high-energy (Section~\ref{Ss:NM}) and low-energy (Section~\ref{Ss:sat})
 ranges were fitted with a prescribed spectral shape, as described below.
Here, we used a modified {Band} function (MBF), which includes an exponential roll-off term, similar to the Ellison-Ramaty
 spectral shape \citep{ellison85}:
\begin{align}
F(R) &= J_1 \left(\frac{R}{1\mathrm{GV}}\right)^{-\gamma_1}\exp \left( -\frac{R}{R_1}\right)  \,\,\,\text{if } R < R_b \label{Eq:MBF_1},\\
F(R) &= J_2 \left(\frac{R}{1\mathrm{GV}}\right)^{-\gamma_2}\exp \left( -\frac{R}{R_2}\right)  \,\,\,\text{if } R \ge R_b,
\label{Eq:MBF_2}
\end{align}
where $F(R)\equiv f(>R)$ is the omnidirectional fluence (in units of cm$^{-2}$) of particles with a rigidity greater than $R$,
 $R$ is expressed in gigavolts, parameters $\gamma_1, \gamma_2, R_1, R_2$, and $J_2$ are defined by fitting, and other
 parameters can be calculated as
\begin{align}
 \label{Eq:J_1}
 R_b &= \gamma_0\cdot {R_0}\\ \nonumber
 J_1 &= J_2 \cdot R_b^{-\gamma_0} \cdot \exp (\gamma_0)\\ \nonumber
 \gamma_0&=\gamma_2-\gamma_1\\ \nonumber
 R_0 &= R_1 \cdot R_2 / (R_2 - R_1).
\end{align}
This function is constructed in such a way that it and its first derivative are continuous, providing a smooth junction between the two parts.
We used a fitting method based on {a three-step Monte Carlo procedure further developed after \citet{usoskin_GLE_20}, as
 illustrated in Figure~\ref{Fig:fit}.
The method includes $L$ Monte Carlo iterations.}

\subsection*{Step 1: Fitting the high-energy (NM) part}

First, we fitted the high-energy part of the spectral shape (Equation~\ref{Eq:MBF_2}) using the data from the NM network (blue crosses in
 Figure~\ref{Fig:fit}) for each event with $M$ spectral points with uncertainties, and $m$ points, corresponding to the upper
 limits of the fluence (vertical blue lines in the figure).
{The uncertainties are both statistical (count-rate statistic) and systematic, related to the `effective-rigidity'
 (or bow-tie) method ones.}
For the $l$-th iteration, we simulated a set of randomised `exact' fluence $F^*(R^*)$ values, {applying the same procedure as
 in \citet{usoskin_GLE_20},} so that for each point $k$ (1..$M$) {the value of $R^*_k$ was randomly and uniformly taken inside
 $R_k$ error bars; the value of $F^*_k(>R)$ was computed using Equation~(4) from \citet{usoskin_GLE_20}, where the scaling factor $K_k$ was
 randomly taken inside the error bars using the uniform distribution, and the GLE {integral} intensity $X_k$ {(in units of \%$\cdot$hr above the GCR background over the entire duration of the event, see \cite{Asvestari2017})}  was randomly taken inside the error bars,
 defined as $\sigma_{X_k} = \max[1; 0.1 \cdot X_k]$ \%$\cdot$hr  using the normal distribution.}
In order to avoid a bias towards more numerous polar NMs during the fitting procedure, we  {considered only} one
 randomly selected point (NM) in each rigidity bin of 0.4 GV widths.
For several weak events (\# 18, 35, 53, 58, 63, 64), the bin width was reduced to 0.1 GV {to keep the number of
 fitted points reasonable.
We checked that reducing the bin size does not lead to any significant bias in the results.}
Then, this set of points $F^*(R^*)$ was fitted by the spectral shape (Equation~\ref{Eq:MBF_2})
 applying a non-linear least-squares method (\textit{\emph{scipy.optimize.curve\_fit}} function in
 Python), and the best-fit parameters $J_2$, $\gamma_2$ and $R_2$ were found based on the minimisation of the logarithmic residual $D$:
\begin{equation}
D \equiv \min \left[  \sum \left(\log(F_{\mathrm{fit}}(R^*)) - \log(F^*(R^*))\right)^2 \right]
\label{eq:D_min}
,\end{equation}
where $F_{\mathrm{fit}}$ is the value computed using the fit-function (Equation~\ref{Eq:MBF_2}) for the rigidity $R^*$.
We additionally checked that the obtained best-fit parameters are physically reasonable, that is, the obtained function does not have
 a positive second derivative anywhere in the high-rigidity ($>$1 GV) range since the differential SEP fluence
  is not expected to increase with $R$ in this range.
This is quantified as the condition that $R_2>0, \gamma_2>0$ or, for $\gamma_2\leq 0$, we required that
$R_2 \left(\sqrt{-\gamma_2} - \gamma_2\right) < 1$.
We also required the formal fit to exceed none of the $m$ upper limits.
Fits that did not satisfy these conditions were discarded, {and the corresponding Monte Carlo simulation was redone
 without counting it in the statistic}.
The obtained best-fit set of parameters ($J_{2,l}$, $\gamma_{2,l,}$ and $R_{2,l})$ was fixed and used in Step 2.

\subsection*{Step 2: Fitting the low-energy part}

The set of $N$ low-energy ($R<$1 GV) fluences $F^*(R^*)$ was obtained in a similar way to Step 1, assuming a fixed value of the
 uncertainty being 10\% (20\% for the pre-GOES/HEPAD era before 1989)\footnote{{This is an ad-hoc order-of-magnitude
 estimate based on the possibility of drawing a smooth curve throughout the experimental points within the error bars, or, on other words,
 keeping the best-fit merit function $\chi_{\rm min}^2$ (Figure~\ref{Fig:chi2}) of the order of unity per degree of freedom.}}
  of the tabulated value and applying {the uniform distribution.}
Only points with $R$~$>$168 MV (corresponding to energy above 15 MeV for protons) were considered, since
 SEPs with the energy $<$15 MeV may have different sources and different spectral parameters \citep[e.g.][]{reames99,cliver_SEP_16}.

Next, each of these $F^*(R^*)$ values was divided by the extrapolated function obtained in Step 1 to form a set of
 values $X(R^*)$:
\begin{equation}
X(R^*) = \frac{F^*(R^*)}  {J_2 \left( {R^*}/1{\rm GV} \right)^{-\gamma_2} \cdot \exp \left( -{R^*}/{R_2}\right) }
.\end{equation}
The obtained set of points $X(R^*)$ was fitted using the same non-linear least-squares method as in Step 1,
 with the following function:
\begin{equation}
x(R) = \left(\frac{R}{\gamma_0 R_0}\right)^{-\gamma_0} \exp \left( -\frac{R}{R_0} + \gamma_0\right)
,\end{equation}
which can be obtained by dividing the low-rigidity part of MBF (Equation~\ref{Eq:MBF_1}) by its high-rigidity part (Equation~\ref{Eq:MBF_2}).
We checked that the obtained fit parameters are {mathematically reasonable}, so the best-fit function does not have a positive derivative
 anywhere, since the integral fluence cannot increase with $R$.
This condition is quantified as $\gamma_1/R+1/R_1>0$ for the rigidity range from 137 MV (equivalent to 10 MeV energy) to $R_\mathrm{b}$.

Thus, for each $l$-th iteration, a set of parameters $J_{2,l}$, $\gamma_{2,l}$, $\gamma_{1,l}$, $R_{2,l}$, $R_{1,l}$
 was calculated for an analysed event.
Then, the formal $\chi^2_l$ value was computed as the merit function between the fitted curve $\mathfrak{F}$ and the data points $F(R)$:
\begin{equation}
\chi^2_l = \sum_{i=1}^{M+N}{\left( {\mathfrak{F}_l(R_{i})-F_i(R_i)\over \sigma_i}\right)^2}.
\label{Eq:chi2}
\end{equation}
This set of parameters and the value of $\chi^2_l$ were recorded for the $l$-th iteration, and then a new ($l$+1)-st iteration started.

This procedure was repeated $L$ = 5000 times, and the best-fit parameters corresponding to the minimum (among all $L$ iterations)
 value of $\chi^2_{\rm min}$ were saved and used in the next step of the procedure.
We also checked points for outliers.
If any data point contributed more than 100 to the total $\chi^2$ of the best-fit option (i.e. laying beyond 10 $\sigma$ 
from the best-fit curve), such a data point was discarded and the fit redone for that event.
Only one outlier of this kind was found -- the point corresponding to the fluence of protons with energy $>$360 MeV for
GLE \#24 from \citet{webber07}.
The set of parameters corresponding to the $\chi^2_{\rm min}$ was selected as the best-fit set and used in Step 3.

\subsection*{Step 3: Evaluation of the uncertainties of the parameters}

\begin{figure*}[t]
\centering
\includegraphics[width=1\textwidth]{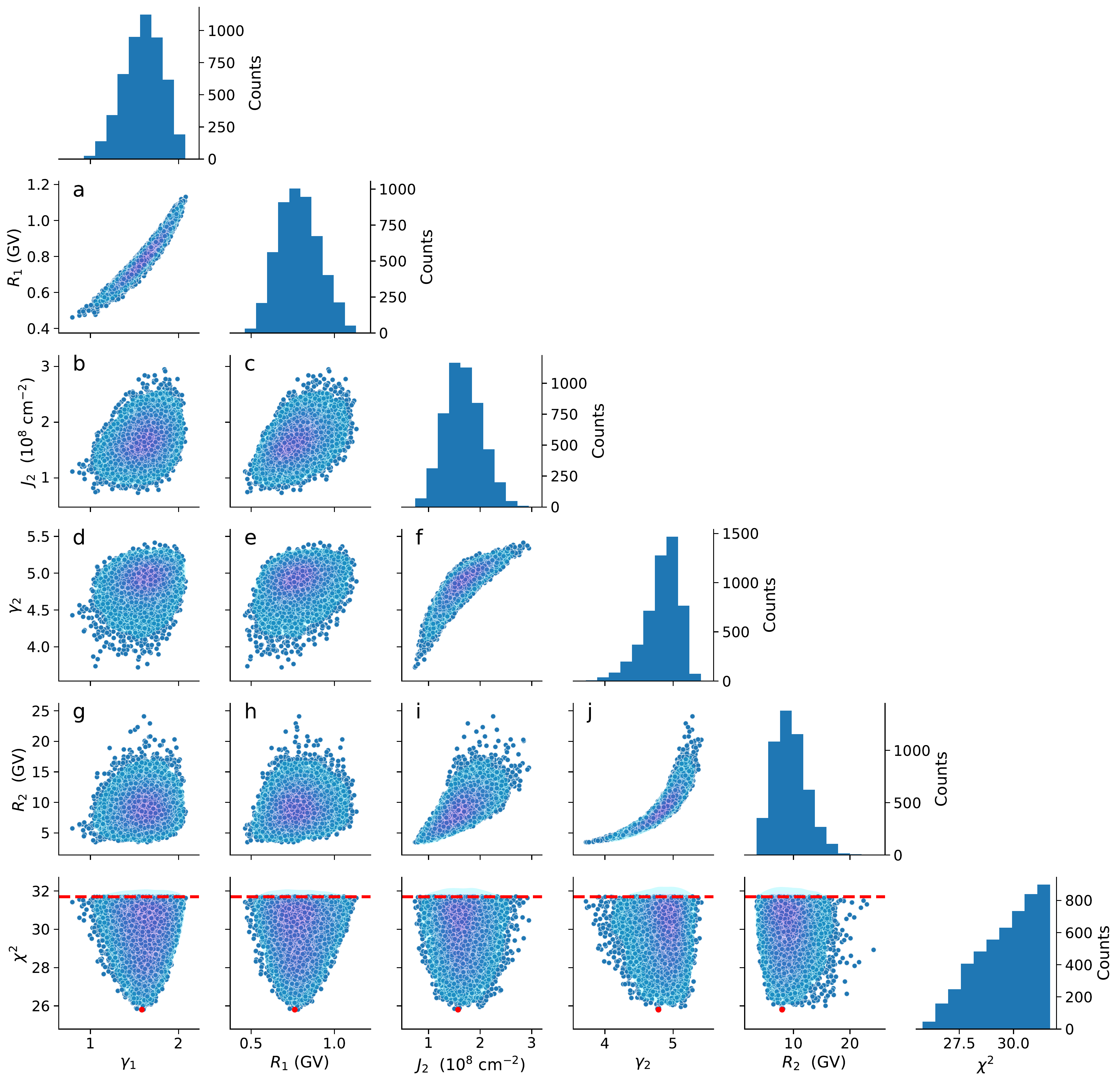}
\caption{
 Distribution of the parameters and their pair-wise correlations for the MBF fitting of the GLE \#5
 obtained for 10000 iterations (see Step 3 in Section~\ref{Sec:fit}).
 The red dots in the bottom panels correspond to the minimum $\chi_{\rm min}^2$, while the horizontal red dashed line
  $\chi^2 = \chi^2_{\rm min}+5.89$ denotes the 68\% confidence interval for the best-fit parameter values. {Colour intensity corresponds to the point density.}
 The diagonal panels depict histograms of the parameter values' distribution.
  }
\label{Fig:chi2}
\end{figure*}

Next, we performed an additional Monte Carlo study of the uncertainties of the obtained best-fit parameters.
The value of each parameter was varied randomly (and independently of each other) so that the new value of a parameter $P$
 (any of the five parameters of the MBF) was taken as
\begin{equation}
P^*=P_0\cdot (1 + r),
\end{equation}
where $P_0$ is the best-fit
 value found in Steps 1 and 2, and $r$ is a normally distributed pseudo-random number with a zero mean and {the standard deviation of 0.5}.
All five parameters of the MBF were simultaneously and independently randomised in this way.
If these parameters were not rejected by the physical-criteria checks (described in Steps 1 and 2), the formal $\chi^2$ value
 (Equation~\ref{Eq:chi2}) was calculated for the corresponding MBF and the data points.
If the obtained  $\chi^2$ value did not exceed $\chi^2_{\rm min}+5.89$, the corresponding set of parameters was recorded as
 being within a 68\% confidence interval (c.i.) for a five-parameter model \citep[e.g. Chapter 15.6 of][]{numrec07}, otherwise it was discarded
 {and the simulation was redone}.
If a value of $\chi^2<\chi^2_{\rm min}$ was obtained during this step, it was assigned as a new $\chi^2_{\rm min}$, and the
 values of the best-fit parameters $P_0$ reset, and Step 3 was restarted anew.
This procedure was repeated 10000 times for each analysed event, which involved (including the discarded iterations)
 $3\cdot 10^8$ iterations for GLE~\#5, shown in Figure~\ref{Fig:chi2}.

Figure~\ref{Fig:chi2} shows an example of such an analysis.
Similar plots were constructed and analysed for all the studied events.
The bottom row of panels depicts the relations between the $\chi^2$ value and the value of each of the five parameters.
Each panel contains 10000 points satisfying the condition of $\chi^2 \leq \chi^2_{\rm min}+5.89$ (see above).
One can see that the relations have the inverted (generally asymmetric) bell-shaped profile, clearly defining the uncertainties for
 each of the parameter.
The corresponding ranges of the MBF fits are shown as grey shaded areas in Figure~\ref{Fig:fit} and Appendix~\ref{App:A}.
However, it would be incorrect to provide 68\% c.i. uncertainties for each parameter independently, since some of them
 are tightly interrelated (e.g. $\gamma_1$ and $R_1$, and $\gamma_2$, $R_2$ and $J_2$), as can be seen in Figure~\ref{Fig:chi2}.
We found that the two parts of the MBF (Equations~\ref{Eq:MBF_1} and \ref{Eq:MBF_2}) can be considered
 independent {(see Figure~\ref{Fig:chi2} (b, c, d, e, g, h)),} but within each part, the parameters' values are highly correlated {(a, f, i, j)}.
As the key parameters, we consider $\gamma_1$ and $\gamma_2$, which are not fully independent in the statistical sense
 (the Pearson's correlation coefficient between them is 0.2), but the power of their common variability is only 4\%, and thus
 their uncertainties can be assumed as roughly independent.
Confidence interval (68\%) uncertainties for $\gamma_1$ and $\gamma_2$ are given in the CDS table.
Uncertainties of other parameters are related to these two key parameters and cannot be considered independent
 (Pearson's correlation coefficients are statistically significant, ranging from 0.86 to 0.96).
Accordingly, the uncertainties of other parameters can be calculated from those of $\gamma_1$ and $\gamma_2$.
The overall quality of the spectral fit can be evaluated via the range $\delta F$
 defined as the average ratio:
\begin{equation}
\delta F = 100\,\% \times \Bigg\langle {F_{\rm up}(R)-F_{\rm low}(R)\over F_{\rm up}(R) + F_{\rm low}(R)}\Bigg\rangle_{R_{\rm s}<R<R_{\rm n}},
\label{Eq:delta}
\end{equation}
where $F_{\rm up}(R)$ and $F_{\rm low}(R)$ are the upper and lower bound values of the MBF for all combinations of the parameters
 within the 68\% c.i. range ($\chi^2_\mathrm{min}+5.89$ -- see Figure~\ref{Fig:chi2}) for each value of rigidity $R$ -; this is depicted
 as the upper and lower envelopes of the
 grey area in Figure~\ref{Fig:fit}, and the averaging is performed in the (logarithmically sampled) rigidity range
 from the lowest rigidity $R_{\rm s}$ for the space-borne data point (typically 230 MV), to the highest rigidity $R_{\rm n}$ of
 non-zero NM data points.
The obtained best-fit MBF parameters {($R_2$ values exceeding 100 GV are given as $+\infty$),} along with the 68\% uncertainties of the key parameters $\gamma_1$ and $\gamma_2,$ as well as
 the values of $\delta F$ for the analysed events are summarised in Table~\ref{T:params} and provided in a readable format in CDS.
CDS also contains numerical values of SEP fluences calculated using the obtained best-fit parameters.
We note that the obtained best-fit spectra are not recommended to be extrapolated to energies below 30 MeV (i.e. beyond the energy
 range used for the fit) as this can lead to unphysical results in the low-energy range.
\begin{center}
\begin{table*}
        \caption{List of the analysed GLE events and best-fit parameters of the modified {Band} function (Equations~\ref{Eq:MBF_1} and \ref{Eq:MBF_2}),
 as well as the 68 \% uncertainty of the fit $\delta F$.
 {$+\infty$ implies the value greater than 100 GV.}
 Parameter $J_1$ can be calculated using Equation~\ref{Eq:J_1}.}
                \label{T:params}
\begin{tabular}{ c c | c  c c c c  c c }
                \hline
GLE \#  &       Date    &                       $\gamma_1$                                              &       $R_1$, GV      &       $J_2$, cm$^{-2}$        &       $\gamma_2$                                              &       $R_2$, GV      &       $R_\mathrm{b}$, GV      &       $\Delta$, \%            \\
\hline                                                                                                                                                                                                                                          
5       &       23-Feb-1956     &$                      1.59                                    $       &$      0.770   $&$     1.63\cdot 10^8    $&$     4.84                                            $&$     8.614   $&$     2.748   $&$     21.0    $       \\
8       &       04-May-1960     &$                      2.85                                    $       &$      -1.276  $&$     9.43\cdot 10^5    $&$     -1.36                                           $&$     0.507   $&$     1.528   $&$     33.8    $       \\
10      &       12-Nov-1960     &$                      3.82                                    $       &$      6.244   $&$     2.71\cdot 10^7    $&$     0.01                                            $&$     0.483   $&$     1.995   $&$     17.0    $       \\
11      &       15-Nov-1960     &$                      2.52                                    $       &$      0.603   $&$     4.20\cdot 10^7    $&$     7.26                                            $&$     34.807  $&$     2.909   $&$     17.5    $       \\
12      &       20-Nov-1960     &$                      1.71                                    $       &$      0.312   $&$     3.73\cdot 10^5    $&$     5.77                                            $&$     +\infty $&$     1.267   $&$     29.0    $       \\
13      &       18-Jul-1961     &$                      1.81                                    $       &$      0.231   $&$     2.60\cdot 10^6    $&$     4.28                                            $&$     0.977   $&$     0.747   $&$     29.8    $       \\
16      &       28-Jan-1967     &$                      1.98                                    $       &$      0.522   $&$     2.12\cdot 10^6    $&$     4.90                                            $&$     4.630   $&$     1.718   $&$     17.5    $       \\
18      &       29-Sep-1968     &$                      -2.68                                   $       &$      0.050   $&$     2.39\cdot 10^4    $&$     4.68                                            $&$     +\infty $&$     0.368   $&$     32.4    $       \\
19      &       18-Nov-1968     &$                      3.97                                    $       &$      0.433   $&$     1.42\cdot 10^5    $&$     5.73                                            $&$     10.705  $&$     0.794   $&$     32.8    $       \\
20      &       25-Feb-1969     &$                      -0.03                                   $       &$      0.124   $&$     1.57\cdot 10^5    $&$     4.47                                            $&$     5.555   $&$     0.571   $&$     27.0    $       \\
21      &       30-Mar-1969     &$                      1.57                                    $       &$      0.540   $&$     9.76\cdot 10^5    $&$     3.00                                            $&$     1.653   $&$     1.147   $&$     19.4    $       \\
22      &       24-Jan-1971     &$                      6.07                                    $       &$      -0.408  $&$     6.72\cdot 10^5    $&$     3.84                                            $&$     1.352   $&$     0.699   $&$     24.5    $       \\
23      &       01-Sep-1971     &$                      2.89                                    $       &$      1.172   $&$     1.19\cdot 10^7    $&$     2.09                                            $&$     0.427   $&$     0.537   $&$     24.1    $       \\
24      &       04-Aug-1972     &$                      2.02                                    $       &$      0.152   $&$     4.35\cdot 10^6    $&$     11.96                                           $&$     +\infty $&$     1.511   $&$     26.7    $       \\
25      &       07-Aug-1972     &$                      1.53                                    $       &$      0.039   $&$     2.60\cdot 10^5    $&$     5.07                                            $&$     +\infty $&$     0.138   $&$     21.5    $       \\
26      &       29-Apr-1973     &$                      -0.18                                   $       &$      0.098   $&$     4.32\cdot 10^4    $&$     3.96                                            $&$     53.011  $&$     0.406   $&$     39.3    $       \\
27      &       30-Apr-1976     &$                      3.06                                    $       &$      0.566   $&$     1.02\cdot 10^5    $&$     5.89                                            $&$     +\infty $&$     1.602   $&$     43.6    $       \\
28      &       19-Sep-1977     &$                      5.49                                    $       &$      -0.288  $&$     2.71\cdot 10^7    $&$     0.03                                            $&$     0.185   $&$     0.615   $&$     35.8    $       \\
29      &       24-Sep-1977     &$                      2.34                                    $       &$      0.558   $&$     2.40\cdot 10^5    $&$     4.78                                            $&$     +\infty $&$     1.362   $&$     26.5    $       \\
30      &       22-Nov-1977     &$                      4.72                                    $       &$      -0.346  $&$     7.03\cdot 10^5    $&$     2.93                                            $&$     1.349   $&$     0.493   $&$     18.9    $       \\
31      &       07-May-1978     &$                      4.27                                    $       &$      -1.585  $&$     3.14\cdot 10^4    $&$     0.02                                            $&$     1.175   $&$     2.868   $&$     23.3    $       \\
32      &       23-Sep-1978     &$                      6.42                                    $       &$      -0.171  $&$     3.50\cdot 10^5    $&$     4.14                                            $&$     3.222   $&$     0.370   $&$     24.3    $       \\
33      &       21-Aug-1979     &$                      5.94                                    $       &$      -1.006  $&$     2.11\cdot 10^5    $&$     0.11                                            $&$     0.383   $&$     1.617   $&$     30.5    $       \\
35      &       10-May-1981     &$                      4.65                                    $       &$      -13.452 $&$     6.45 \cdot 10^{11}   $&$     -15.17                                          $&$     0.059   $&$     1.164   $&$     49.1    $       \\
36      &       12-Oct-1981     &$                      7.40                                    $       &$      -0.158  $&$     5.59\cdot 10^5    $&$     3.35                                            $&$     2.857   $&$     0.606   $&$     28.3    $       \\
37      &       26-Nov-1982     &$                      4.90                                    $       &$      -0.976  $&$     6.99\cdot 10^4    $&$     1.67                                            $&$     1.287   $&$     1.793   $&$     26.8    $       \\
38      &       07-Dec-1982     &$                      5.96                                    $       &$      -0.250  $&$     8.58\cdot 10^5    $&$     2.05                                            $&$     0.922   $&$     0.769   $&$     27.3    $       \\
39      &       16-Feb-1984     &$                      3.76                                    $       &$      -29.660 $&$     1.30\cdot 10^5    $&$     0.21                                            $&$     0.715   $&$     2.479   $&$     31.6    $       \\
40      &       25-Jul-1989     &$                      -0.24                                   $       &$      0.154   $&$     7.33\cdot 10^4    $&$     6.18                                            $&$     +\infty $&$     0.989   $&$     26.0    $       \\
41      &       15-Aug-1989     &$                      2.10                                    $       &$      0.315   $&$     1.65\cdot 10^6    $&$     4.25                                            $&$     1.375   $&$     0.879   $&$     15.0    $       \\
42      &       29-Sep-1989     &$                      -1.95                                   $       &$      0.069   $&$     9.89\cdot 10^6    $&$     3.39                                            $&$     5.935   $&$     0.373   $&$     9.6     $       \\
43      &       19-Oct-1989     &$                      -1.21                                   $       &$      0.109   $&$     6.59\cdot 10^6    $&$     4.44                                            $&$     2.960   $&$     0.639   $&$     13.5    $       \\
44      &       22-Oct-1989     &$                      0.35                                    $       &$      0.122   $&$     1.84\cdot 10^6    $&$     6.07                                            $&$     5.807   $&$     0.713   $&$     15.6    $       \\
45      &       24-Oct-1989     &$                      -1.77                                   $       &$      0.077   $&$     1.19\cdot 10^7    $&$     2.95                                            $&$     1.131   $&$     0.390   $&$     13.0    $       \\
46      &       15-Nov-1989     &$                      -1.16                                   $       &$      0.117   $&$     5.58\cdot 10^4    $&$     4.97                                            $&$     +\infty $&$     0.717   $&$     20.4    $       \\
47      &       21-May-1990     &$                      -1.55                                   $       &$      0.088   $&$     2.85\cdot 10^5    $&$     4.06                                            $&$     +\infty $&$     0.494   $&$     12.4    $       \\
48      &       24-May-1990     &$                      -0.95                                   $       &$      0.137   $&$     6.14\cdot 10^5    $&$     4.08                                            $&$     7.501   $&$     0.702   $&$     15.9    $       \\
49      &       26-May-1990     &$                      -1.03                                   $       &$      0.131   $&$     2.80\cdot 10^5    $&$     4.51                                            $&$     +\infty $&$     0.726   $&$     14.8    $       \\
50      &       28-May-1990     &$                      -0.80                                   $       &$      0.142   $&$     1.67\cdot 10^5    $&$     5.51                                            $&$     +\infty $&$     0.900   $&$     21.1    $       \\
51      &       11-Jun-1991     &$                      -2.31                                   $       &$      0.059   $&$     1.55\cdot 10^5    $&$     4.87                                            $&$     +\infty $&$     0.424   $&$     14.2    $       \\
52      &       15-Jun-1991     &$                      1.13                                    $       &$      0.200   $&$     8.92\cdot 10^5    $&$     5.02                                            $&$     2.883   $&$     0.836   $&$     15.4    $       \\
53      &       25-Jun-1992     &$                      2.12                                    $       &$      0.297   $&$     8.89\cdot 10^4    $&$     5.28                                            $&$     +\infty $&$     0.939   $&$     22.8    $       \\
55      &       06-Nov-1997     &$                      0.75                                    $       &$      0.165   $&$     9.66\cdot 10^5    $&$     4.23                                            $&$     3.077   $&$     0.607   $&$     15.6    $       \\
56      &       02-May-1998     &$                      1.57                                    $       &$      0.221   $&$     6.26\cdot 10^4    $&$     5.01                                            $&$     +\infty $&$     0.760   $&$     25.9    $       \\
58      &       24-Aug-1998     &$                      4.60                                    $       &$      -0.277  $&$     1.55\cdot 10^6    $&$     1.65                                            $&$     0.300   $&$     0.425   $&$     27.5    $       \\
59      &       14-Jul-2000     &$                      2.80                                    $       &$      0.276   $&$     3.46\cdot 10^6    $&$     5.85                                            $&$     2.191   $&$     0.963   $&$     17.7    $       \\
60      &       15-Apr-2001     &$                      1.02                                    $       &$      0.290   $&$     2.90\cdot 10^6    $&$     4.57                                            $&$     3.321   $&$     1.128   $&$     15.7    $       \\
61      &       18-Apr-2001     &$                      2.37                                    $       &$      0.547   $&$     3.68\cdot 10^5    $&$     4.01                                            $&$     5.172   $&$     1.003   $&$     20.5    $       \\
62      &       04-Nov-2001     &$                      3.70                                    $       &$      0.311   $&$     3.08\cdot 10^5    $&$     6.63                                            $&$     +\infty $&$     0.911   $&$     18.6    $       \\
63      &       26-Dec-2001     &$                      2.45                                    $       &$      0.245   $&$     5.79\cdot 10^5    $&$     4.11                                            $&$     0.673   $&$     0.640   $&$     20.7    $       \\
64      &       24-Aug-2002     &$                      2.09                                    $       &$      0.281   $&$     1.69\cdot 10^5    $&$     7.00                                            $&$     +\infty $&$     1.380   $&$     33.9    $       \\
65      &       28-Oct-2003     &$                      6.49                                    $       &$      -0.525  $&$     5.31\cdot 10^6    $&$     0.01                                            $&$     0.447   $&$     1.565   $&$     22.0    $       \\
66      &       29-Oct-2003     &$                      1.56                                    $       &$      0.188   $&$     5.19\cdot 10^5    $&$     6.85                                            $&$     +\infty $&$     0.995   $&$     23.9    $       \\
67      &       02-Nov-2003     &$                      3.62                                    $       &$      0.541   $&$     1.81\cdot 10^5    $&$     5.40                                            $&$     +\infty $&$     0.963   $&$     20.6    $       \\
69      &       20-Jan-2005     &$                      1.16                                    $       &$      0.295   $&$     8.03\cdot 10^6    $&$     4.73                                            $&$     2.392   $&$     1.201   $&$     16.3    $       \\
70      &       13-Dec-2006     &$                      0.18                                    $       &$      0.114   $&$     1.46\cdot 10^6    $&$     3.77                                            $&$     1.449   $&$     0.444   $&$     14.5    $       \\
71      &       16-May-2012     &$                      2.23                                    $       &$      0.372   $&$     2.60\cdot 10^5    $&$     7.03                                            $&$     24.488  $&$     1.813   $&$     10.5    $       \\
72      &       10-Sep-2017     &$                      1.46                                    $       &$      0.164   $&$     4.59\cdot 10^5    $&$     6.31                                            $&$     +\infty $&$     0.795   $&$     20.9    $       \\
                \hline
\end{tabular}
        \end{table*}
\end{center}

\section{Conclusions}

In this work, event-integrated fluences of solar energetic particles were re-evaluated for most
 major SEP events (GLEs) using updated low-energy flux estimates, greatly improved high-energy flux data,
 and the newly developed reconstruction methods.
The earlier estimates \citep{tylka09,raukunen18} were essentially revisited here, providing an accurate
 parametrisation of the rigidity spectra, with the spectral parameters being tabulated in Table~\ref{T:params}
 and provided in CDS.
In particular, {it was shown earlier \citep[Fig.~6]{usoskin_GLE_20} that the {Band} function spectral
 shape can lead to a significant overestimate of the high-energy tail of the spectrum}, and a strong roll-off (assumed to be exponential here)
 is required.
Accordingly, for the parametrisation of the spectral fluence shape, we propose a modified {Band} function
 (Equations~\ref{Eq:MBF_1}\,--\,\ref{Eq:MBF_2}), which is a combination of the standard {Band} and a modified
 Ellison-Ramaty functions.
The spectral fluences evaluated with the revisited datasets and a new method
 form a solid basis for more precise studies of the physics of solar eruptive events and {the transport} of energetic
 particles in the interstellar medium.

The new spectral fluences will be also useful in various applications of SEP {influence on} terrestrial effects
 such as cosmic-ray-induced ionisation \citep[e.g.,][]{jackman08,usoskin_ACP_11,duderstadt16},
 radiation hazards \citep[e.g.,][]{feynman93,Jiggens2014,mishev15,raukunen18},
 or cosmogenic isotope production \citep{webber07,usoskin_F200_14,mekhaldi15}.
The latter is of particular importance for studies of the reference SEP events \citep{cliver14,usoskin_1956_20}, and,
 accordingly, to a more precise assessment of historical extreme solar particle storms \citep{miyake19}.

\begin{acknowledgements}
Detrended data of GLE recorded by NMs were obtained from the International GLE database https://gle.oulu.fi.
PIs and teams of all the ground-based neutron monitors and space-borne experiments whose data were used here are gratefully acknowledged.
This work was partially supported by the Academy of Finland (project No. 321882 ESPERA). {Collection of the data from cosmic-ray/ionospheric experiments before 1989 and the creation of full dataset used in this work was supported by the Russian Science Foundation project no. 20-72-10170. Development of the fluence fitting procedure was supported by the Russian Science Foundation project no. 20-67-46016.}
The work in the University of Turku was performed in the framework of the Finnish Centre of Excellence in Research of
 Sustainable Space funded by the Academy of Finland (grant no. 312357).
The authors benefited from discussions within the ISSI International Team work (HEROIC team) and ISWAT-COSPAR S1-02 team. %\textbf{The authors are grateful to Patrick K\"{u}hl for his careful review of this work}
\end{acknowledgements}

%% ------------------------------------------------------------------------ %%
%% References and Citations

%%%%%%%%%%%%%%%%%%%%%%%%%%%%%%%%%%%%%%%%%%%%%%%
%
% \bibliography{<name of your .bib file>} don't specify the file extension
%
% don't specify bibliographystyle
%%%%%%%%%%%%%%%%%%%%%%%%%%%%%%%%%%%%%%%%%%%%%%%
\bibliographystyle{aa}
%\bibliography{bibliography.bib}

\bibliography{usoskin_all,library_Sergey}

\begin{thebibliography}{56}
\expandafter\ifx\csname natexlab\endcsname\relax\def\natexlab#1{#1}\fi

\bibitem[{{Adriani} {et~al.}(2014){Adriani}, {Barbarino}, {Bazilevskaya},
  {Bellotti}, {Boezio}, {Bogomolov}, {Bongi}, {Bonvicini}, {Bottai}, {Bruno},
  {Cafagna}, {Campana}, {Carbone}, {Carlson}, {Casolino}, {Castellini}, {De
  Pascale}, {De Santis}, {De Simone}, {Di Felice}, {Formato}, {Galper},
  {Giaccari}, {Karelin}, {Kheymits}, {Koldashov}, {Koldobskiy}, {Krut`kov},
  {Kvashnin}, {Leonov}, {Malakhov}, {Marcelli}, {Martucci}, {Mayorov}, {Menn},
  {Mikhailov}, {Mocchiutti}, {Monaco}, {Mori}, {Munini}, {Nikonov}, {Osteria},
  {Papini}, {Pearce}, {Picozza}, {Pizzolotto}, {Ricci}, {Ricciarini},
  {Rossetto}, {Sarkar}, {Simon}, {Sparvoli}, {Spillantini}, {Stozhkov},
  {Vacchi}, {Vannuccini}, {Vasilyev}, {Voronov}, {Wu}, {Yurkin}, {Zampa},
  {Zampa}, \& {Zverev}}]{adriani14}
{Adriani}, O., {Barbarino}, G.~C., {Bazilevskaya}, G.~A., {et~al.} 2014, Phys.
  Rep., 544, 323

\bibitem[{{Aguilar} {et~al.}(2018){Aguilar}, {Ali Cavasonza}, {Alpat},
  {Ambrosi}, {Arruda}, {Attig}, {Aupetit}, {Azzarello}, {Bachlechner}, {Barao},
  {Barrau}, {Barrin}, {Bartoloni}, {Basara}, {Ba{\textcommabelow
  s}e{\v{g}}mez-du Pree}, {Battarbee}, {Battiston}, {Becker}, {Behlmann},
  {Beischer}, {Berdugo}, {Bertucci}, {Bindel}, {Bindi}, {de Boer}, {Bollweg},
  {Bonnivard}, {Borgia}, {Boschini}, {Bourquin}, {Bueno}, {Burger}, {Cadoux},
  {Cai}, {Capell}, {Caroff}, {Casaus}, {Castellini}, {Cervelli}, {Chae},
  {Chang}, {Chen}, {Chen}, {Chen}, {Chen}, {Cheng}, {Chou}, {Choumilov},
  {Choutko}, {Chung}, {Clark}, {Clavero}, {Coignet}, {Consolandi}, {Contin},
  {Corti}, {Creus}, {Crispoltoni}, {Cui}, {Dadzie}, {Dai}, {Datta}, {Delgado},
  {Della Torre}, {Demirk{\"o}z}, {Derome}, {Di Falco}, {Dimiccoli},
  {D{\'\i}az}, {von Doetinchem}, {Dong}, {Donnini}, {Duranti}, {D'Urso},
  {Egorov}, {Eline}, {Eronen}, {Feng}, {Fiandrini}, {Fisher}, {Formato},
  {Galaktionov}, {Gallucci}, {Garc{\'\i}a-L{\'o}pez}, {Gargiulo}, {Gast},
  {Gebauer}, {Gervasi}, {Ghelfi}, {Giovacchini}, {G{\'o}mez-Coral}, {Gong},
  {Goy}, {Grabski}, {Grandi}, {Graziani}, {Guo}, {Haino}, {Han}, {He}, {Heil},
  {Hoffman}, {Hsieh}, {Huang}, {Huang}, {Huh}, {Incagli}, {Ionica}, {Jang},
  {Jia}, {Jinchi}, {Kang}, {Kanishev}, {Khiali}, {Kim}, {Kim}, {Kirn}, {Konak},
  {Kounina}, {Kounine}, {Koutsenko}, {Kulemzin}, {La Vacca}, {Laudi},
  {Laurenti}, {Lazzizzera}, {Lebedev}, {Lee}, {Lee}, {Leluc}, {Li}, {Li}, {Li},
  {Li}, {Li}, {Li}, {Light}, {Lim}, {Lin}, {Lipari}, {Lippert}, {Liu}, {Liu},
  {Lordello}, {Lu}, {Lu}, {Luebelsmeyer}, {Luo}, {Luo}, {Luo}, {Lyu},
  {Machate}, {Ma{\~n}{\'a}}, {Mar{\'\i}n}, {Martin}, {Mart{\'\i}nez}, {Masi},
  {Maurin}, {Menchaca-Rocha}, {Meng}, {Mikuni}, {Mo}, {Mott}, {Nelson}, {Ni},
  {Nikonov}, {Nozzoli}, {Oliva}, {Orcinha}, {Palermo}, {Palmonari},
  {Palomares}, {Paniccia}, {Pauluzzi}, {Pensotti}, {Perrina}, {Phan},
  {Picot-Clemente}, {Pilo}, {Pizzolotto}, {Plyaskin}, {Pohl}, {Poireau},
  {Popkow}, {Quadrani}, {Qi}, {Qin}, {Qu}, {R{\"a}ih{\"a}}, {Rancoita},
  {Rapin}, {Ricol}, {Rosier-Lees}, {Rozhkov}, {Rozza}, {Sagdeev}, {Schael},
  {Schmidt}, {Schulz von Dratzig}, {Schwering}, {Seo}, {Shan}, {Shi},
  {Siedenburg}, {Son}, {Song}, {Tacconi}, {Tang}, {Tang}, {Tescaro}, {Ting},
  {Ting}, {Tomassetti}, {Torsti}, {T{\"u}rko{\v{g}}lu}, {Urban}, {Vagelli},
  {Valente}, {Valtonen}, {V{\'a}zquez Acosta}, {Vecchi}, {Velasco}, {Vialle},
  {Wang}, {Wang}, {Wang}, {Wang}, {Wang}, {Wang}, {Wei}, {Weng}, {Whitman},
  {Wu}, {Wu}, {Xiong}, {Xu}, {Yan}, {Yang}, {Yang}, {Yang}, {Yi}, {Yu}, {Yu},
  {Zannoni}, {Zeissler}, {Zhang}, {Zhang}, {Zhang}, {Zhang}, {Zhang}, {Zhang},
  {Zheng}, {Zhuang}, {Zhukov}, {Zichichi}, {Zimmermann}, {Zuccon}, \& {AMS
  Collaboration}}]{aguilar_AMS_18}
{Aguilar}, M., {Ali Cavasonza}, L., {Alpat}, B., {et~al.} 2018, Phys. Rev.
  Lett., 121, 051101

\bibitem[{Anastasiadis {et~al.}(2019)Anastasiadis, Lario, Papaioannou,
  Kouloumvakos, \& Vourlidas}]{Anastasiadis2019}
Anastasiadis, A., Lario, D., Papaioannou, A., Kouloumvakos, A., \& Vourlidas,
  A. 2019, Philos. T. R. Soc. A, 377, 20180100

\bibitem[{Asvestari {et~al.}(2017)Asvestari, Gil, Kovaltsov, \&
  Usoskin}]{Asvestari2017}
Asvestari, E., Gil, A., Kovaltsov, G.~A., \& Usoskin, I.~G. 2017, J.
  Geophys. Res. (Space Phys.), 122, 9790

\bibitem[{{Band} {et~al.}(1993){Band}, {Matteson}, {Ford}, {Schaefer},
  {Palmer}, {Teegarden}, {Cline}, {Briggs}, {Paciesas}, {Pendleton}, {Fishman},
  {Kouveliotou}, {Meegan}, {Wilson}, \& {Lestrade}}]{band93}
{Band}, D., {Matteson}, J., {Ford}, L., {et~al.} 1993, Astrophys. J., 413, 281

\bibitem[{{Bazilevskaya} {et~al.}(2014){Bazilevskaya}, {Cliver}, {Kovaltsov},
  {Ling}, {Shea}, {Smart}, \& {Usoskin}}]{bazilevskaya14}
{Bazilevskaya}, G.~A., {Cliver}, E.~W., {Kovaltsov}, G.~A., {et~al.} 2014,
  Space Sci. Rev., 186, 409

\bibitem[{{Bindi}(2017)}]{bindi17}
{Bindi}, V. 2017, Adv. Space Res., 60, 753

\bibitem[{{Bruno} {et~al.}(2018){Bruno}, {Bazilevskaya}, {Boezio}, {Christian},
  {de Nolfo}, {Martucci}, {Merge'}, {Mikhailov}, {Munini}, {Richardson},
  {Ryan}, {Stochaj}, {Adriani}, {Barbarino}, {Bellotti}, {Bogomolov}, {Bongi},
  {Bonvicini}, {Bottai}, {Cafagna}, {Campana}, {Carlson}, {Casolino},
  {Castellini}, {De Santis}, {Di Felice}, {Galper}, {Karelin}, {Koldashov},
  {Koldobskiy}, {Krutkov}, {Kvashnin}, {Leonov}, {Malakhov}, {Marcelli},
  {Mayorov}, {Menn}, {Mocchiutti}, {Monaco}, {Mori}, {Osteria}, {Panico},
  {Papini}, {Pearce}, {Picozza}, {Ricci}, {Ricciarini}, {Simon}, {Sparvoli},
  {Spillantini}, {Stozhkov}, {Vacchi}, {Vannuccini}, {Vasilyev}, {Voronov},
  {Yurkin}, {Zampa}, \& {Zampa}}]{bruno18}
{Bruno}, A., {Bazilevskaya}, G.~A., {Boezio}, M., {et~al.} 2018, Astrophys. J.,
  862, 97

\bibitem[{Clem \& Dorman(2000)}]{clem00}
Clem, J. \& Dorman, L. 2000, Space Sci. Rev., 93, 335

\bibitem[{{Cliver}(2016)}]{cliver_SEP_16}
{Cliver}, E.~W. 2016, Astrophys. J., 832, 128

\bibitem[{{Cliver} {et~al.}(2020){Cliver}, {Hayakawa}, {Love}, \&
  {Neidig}}]{cliver20}
{Cliver}, E.~W., {Hayakawa}, H., {Love}, J.~J., \& {Neidig}, D.~F. 2020,
  Astrophys. J., 903, 41

\bibitem[{{Cliver} {et~al.}(2014){Cliver}, {Tylka}, {Dietrich}, \&
  {Ling}}]{cliver14}
{Cliver}, E.~W., {Tylka}, A.~J., {Dietrich}, W.~F., \& {Ling}, A.~G. 2014,
  Astrophys. J., 781, 32

\bibitem[{{Desai} \& {Giacalone}(2016)}]{desai_LR_16}
{Desai}, M. \& {Giacalone}, J. 2016, Liv. Rev. Solar Phys., 13, 3

\bibitem[{{Duderstadt} {et~al.}(2016){Duderstadt}, {Dibb}, {Schwadron},
  {Spence}, {Solomon}, {Yudin}, {Jackman}, \& {Randall}}]{duderstadt16}
{Duderstadt}, K.~A., {Dibb}, J.~E., {Schwadron}, N.~A., {et~al.} 2016, J.
  Geophys. Res. (Atmos.), 121, 2994

\bibitem[{{Ellison} \& {Ramaty}(1985)}]{ellison85}
{Ellison}, D.~C. \& {Ramaty}, R. 1985, Astrophys. J., 298, 400

\bibitem[{Feynman \& Gabriel(1990)}]{feynman90}
Feynman, J. \& Gabriel, S. 1990, Solar Phys., 127, 393

\bibitem[{{Feynman} {et~al.}(1993){Feynman}, {Spitale}, {Wang}, \&
  {Gabriel}}]{feynman93}
{Feynman}, J., {Spitale}, G., {Wang}, J., \& {Gabriel}, S. 1993, J. Geophys.
  Res., 98, 13

\bibitem[{{Forbush}(1946)}]{forbush46}
{Forbush}, S.~E. 1946, Phys. Rev., 70, 771

\bibitem[{Goswami {et~al.}(1988)Goswami, McGuire, Reedy, Lal, \&
  Jha}]{goswami88}
Goswami, J., McGuire, R., Reedy, R., Lal, D., \& Jha, R. 1988, J. Geophys.
  Res., 93, 7195

\bibitem[{Herbst {et~al.}(2019)Herbst, Grenfell, Sinnhuber, Rauer, Heber,
  Banjac, Scheucher, Schmidt, Gebauer, Lehmann, \& Schreier}]{Herbst2019}
Herbst, K., Grenfell, J.~L., Sinnhuber, M., {et~al.} 2019, Astron.
  Astrophys., 631, A101

\bibitem[{{Jackman} {et~al.}(2008){Jackman}, {Marsh}, {Vitt}, {Garcia},
  {Fleming}, {Labow}, {Randall}, {L{\'o}pez-Puertas}, {Funke}, {von Clarmann},
  \& {Stiller}}]{jackman08}
{Jackman}, C.~H., {Marsh}, D.~R., {Vitt}, F.~M., {et~al.} 2008, Atmos. Chem.
  Phys., 8, 765

\bibitem[{Jiggens {et~al.}(2014)Jiggens, Chavy-Macdonald, Santin, Menicucci,
  Evans, \& Hilgers}]{Jiggens2014}
Jiggens, P., Chavy-Macdonald, M.-A., Santin, G., {et~al.} 2014, J.
  Space Weather Space Clim., 4, A20

\bibitem[{{Jun} {et~al.}(2007){Jun}, {Swimm}, {Ruzmaikin}, {Feynman}, {Tylka},
  \& {Dietrich}}]{jun07}
{Jun}, I., {Swimm}, R.~T., {Ruzmaikin}, A., {et~al.} 2007, Adv. Space Res., 40,
  304

\bibitem[{King(1974)}]{king74}
King, J. 1974, J. Spacecraft Rockets, 11, 401

\bibitem[{{Klein} \& {Dalla}(2017)}]{klein17}
{Klein}, K.-L. \& {Dalla}, S. 2017, Space Sci. Rev., 212, 1107

\bibitem[{{Kocharov} {et~al.}(2017){Kocharov}, {Pohjolainen}, {Mishev},
  {Reiner}, {Lee}, {Laitinen}, {Didkovsky}, {Pizzo}, {Kim}, {Klassen},
  {Karlicky}, {Cho}, {Gary}, {Usoskin}, {Valtonen}, \& {Vainio}}]{kocharov17}
{Kocharov}, L., {Pohjolainen}, S., {Mishev}, A., {et~al.} 2017, Astrophys. J.,
  839, 79

\bibitem[{{Koldobskiy} {et~al.}(2019){Koldobskiy}, {Kovaltsov}, {Mishev}, \&
  {Usoskin}}]{koldobsky_Reff_19}
{Koldobskiy}, S.~A., {Kovaltsov}, G.~A., {Mishev}, A.~L., \& {Usoskin}, I.~G.
  2019, Solar Phys., 294, 94

\bibitem[{{Koldobskiy} {et~al.}(2018){Koldobskiy}, {Kovaltsov}, \&
  {Usoskin}}]{koldobsky_Eff_2018}
{Koldobskiy}, S.~A., {Kovaltsov}, G.~A., \& {Usoskin}, I.~G. 2018, Solar Phys.,
  293, 110

\bibitem[{{Kong} {et~al.}(2017){Kong}, {Guo}, {Giacalone}, {Li}, \&
  {Chen}}]{kong17}
{Kong}, X., {Guo}, F., {Giacalone}, J., {Li}, H., \& {Chen}, Y. 2017,
  Astrophys. J., 851, 38

\bibitem[{{Kouloumvakos} {et~al.}(2015){Kouloumvakos}, {Nindos}, {Valtonen},
  {Alissand rakis}, {Malandraki}, {Tsitsipis}, {Kontogeorgos}, {Moussas}, \&
  {Hillaris}}]{kouloumvakos15}
{Kouloumvakos}, A., {Nindos}, A., {Valtonen}, E., {et~al.} 2015, Astron.
  Astrophys, 580, A80

\bibitem[{{Kovaltsov} {et~al.}(2014){Kovaltsov}, {Usoskin}, {Cliver},
  {Dietrich}, \& {Tylka}}]{usoskin_F200_14}
{Kovaltsov}, G.~A., {Usoskin}, I.~G., {Cliver}, E.~W., {Dietrich}, W.~F., \&
  {Tylka}, A.~J. 2014, Solar Phys., 289, 4691

\bibitem[{Mekhaldi {et~al.}(2015)Mekhaldi, Muscheler, Adolphi, Aldahan, Beer,
  McConnell, Possnert, Sigl, Svensson, Synal, Welten, \& Woodruff}]{mekhaldi15}
Mekhaldi, F., Muscheler, R., Adolphi, F., {et~al.} 2015, Nature Comm., 6, 8611

\bibitem[{{Mishev} {et~al.}(2015){Mishev}, {Adibpour}, {Usoskin}, \&
  {Felsberger}}]{mishev15}
{Mishev}, A., {Adibpour}, F., {Usoskin}, I., \& {Felsberger}, E. 2015, Adv.
  Space Res., 55, 354

\bibitem[{{Mishev} {et~al.}(2020){Mishev}, {Koldobskiy}, {Kovaltsov}, {Gil}, \&
  {Usoskin}}]{mishev20}
{Mishev}, A., {Koldobskiy}, S., {Kovaltsov}, G., {Gil}, A., \& {Usoskin}, I.
  2020, J. Geophys. Res. (Space Phys.), 125, {e2019JA027433}

\bibitem[{{Mishev} {et~al.}(2013){Mishev}, {Usoskin}, \&
  {Kovaltsov}}]{mishev13}
{Mishev}, A., {Usoskin}, I., \& {Kovaltsov}, G. 2013, J. Geophys. Res. (Space
  Phys.), 118, 2783

\bibitem[{{Mishev} {et~al.}(2018){Mishev}, {Usoskin}, {Raukunen}, {Paassilta},
  {Valtonen}, {Kocharov}, \& {Vainio}}]{mishev18}
{Mishev}, A., {Usoskin}, I., {Raukunen}, O., {et~al.} 2018, Solar Phys., 293,
  136

\bibitem[{Miyake {et~al.}(2019)Miyake, Usoskin, \& Poluianov}]{miyake19}
Miyake, F., Usoskin, I., \& Poluianov, S., eds. 2019, Extreme Solar Particle
  Storms: The Hostile Sun (Bristol, UK: IOP Publishing)

\bibitem[{Onsager {et~al.}(1996)Onsager, Grubb, Kunches, Matheson, Speich,
  Zwickl, \& Sauer}]{Onsager1996}
Onsager, T., Grubb, R., Kunches, J., {et~al.} 1996, in GOES-8 and Beyond, ed.
  E.~R. Washwell, Vol. 2812 (SPIE), 281--290

\bibitem[{{Plainaki} {et~al.}(2014){Plainaki}, {Mavromichalaki}, {Laurenza},
  {Gerontidou}, {Kanellakopoulos}, \& {Storini}}]{Plainaki14}
{Plainaki}, C., {Mavromichalaki}, H., {Laurenza}, M., {et~al.} 2014, Astrophys.
  J., 785, 160

\bibitem[{Press {et~al.}(2007)Press, Teukolsky, Vetterling, \&
  Flannery}]{numrec07}
Press, W., Teukolsky, S., Vetterling, W., \& Flannery, B. 2007, Numerical
  Recipes 3rd Edition: The Art of Scientific Computing, 3rd edn. (USA:
  Cambridge University Press)

\bibitem[{{Raukunen} {et~al.}(2020){Raukunen}, {Paassilta}, {Vainio},
  {Rodriguez}, {Eronen}, {Crosby}, {Dierckxsens}, {Jiggens}, {Heynderickx}, \&
  {Sandberg}}]{raukunen20}
{Raukunen}, O., {Paassilta}, M., {Vainio}, R., {et~al.} 2020, J. Space Weather
  Space Clim., 10, 24

\bibitem[{{Raukunen} {et~al.}(2018){Raukunen}, {Vainio}, {Tylka}, {Dietrich},
  {Jiggens}, {Heynderickx}, {Dierckxsens}, {Crosby}, {Ganse}, \&
  {Siipola}}]{raukunen18}
{Raukunen}, O., {Vainio}, R., {Tylka}, A.~J., {et~al.} 2018, J. Space Weather
  Space Clim., 8, {A04}

\bibitem[{{Reames}(1999)}]{reames99}
{Reames}, D.~V. 1999, Space Sci. Rev., 90, 413

\bibitem[{Reedy(1977)}]{reedy77}
Reedy, R. 1977, in Lunar and Planetary Science VIII, ed. R.~Merril (Houston,
  U.S.A.: Lunar and Planetary Institute), 825--839

\bibitem[{Reeves {et~al.}(1992)Reeves, Cayton, Gary, \& Belian}]{reeves92}
Reeves, G., Cayton, T., Gary, S., \& Belian, R. 1992, J. Geophys. Res., 97,
  6219

\bibitem[{Sellers \& Hanser(1996)}]{Sellers1996}
Sellers, F.~B. \& Hanser, F.~A. 1996, in GOES-8 and Beyond, ed. E.~R. Washwell,
  Vol. 2812 (SPIE), 353--364

\bibitem[{{Shea} \& {Smart}(2012)}]{shea12}
{Shea}, M.~A. \& {Smart}, D.~F. 2012, Space Sci. Rev., 171, 161

\bibitem[{Tylka \& Dietrich(2009)}]{tylka09}
Tylka, A. \& Dietrich, W. 2009, in 31th International Cosmic Ray Conference
  (Lod\'z, Poland: Universal Academy Press), icrc0273

\bibitem[{Tylka {et~al.}(1997)Tylka, Dietrich, \& Boberg}]{tylka97}
Tylka, A., Dietrich, W., \& Boberg, P. 1997, IEEE Trans. Nucl. Sci., 44, 2140

\bibitem[{{Usoskin} {et~al.}(2020{\natexlab{a}}){Usoskin}, {Koldobskiy},
  {Kovaltsov}, {Rozanov}, {Sukhodolov}, {Mishev}, \&
  {Mironova}}]{usoskin_1956_20}
{Usoskin}, I., {Koldobskiy}, S., {Kovaltsov}, G., {et~al.} 2020{\natexlab{a}},
  J. Geophys. Res. (Space Phys.), 125, e27921

\bibitem[{{Usoskin} {et~al.}(2020{\natexlab{b}}){Usoskin}, {Koldobskiy},
  {Kovaltsov}, {Gil}, {Usoskina}, {Willamo}, \& {Ibragimov}}]{usoskin_GLE_20}
{Usoskin}, I., {Koldobskiy}, S., {Kovaltsov}, G.~A., {et~al.}
  2020{\natexlab{b}}, Astron. Astrophys., 640, A17

\bibitem[{{Usoskin} {et~al.}(2011){Usoskin}, {Kovaltsov}, {Mironova}, {Tylka},
  \& {Dietrich}}]{usoskin_ACP_11}
{Usoskin}, I.~G., {Kovaltsov}, G.~A., {Mironova}, I.~A., {Tylka}, A.~J., \&
  {Dietrich}, W.~F. 2011, Atmos. Chem. Phys., 11, 1979

\bibitem[{{Vainio} \& {Afanasiev}(2018)}]{vainio18}
{Vainio}, R. \& {Afanasiev}, A. 2018, Astrophys. Space Sci. Lib., Vol. 444,
  {Particle Acceleration Mechanisms}, ed. O.~{Malandraki} \& N.~{Crosby},
  {45--61}

\bibitem[{{Vainio} {et~al.}(2009){Vainio}, {Desorgher}, {Heynderickx},
  {Storini}, {Fl{\"u}ckiger}, {Horne}, {Kovaltsov}, {Kudela}, {Laurenza},
  {McKenna-Lawlor}, {Rothkaehl}, \& {Usoskin}}]{vainio09}
{Vainio}, R., {Desorgher}, L., {Heynderickx}, D., {et~al.} 2009, Space Sci.
  Rev., 147, 187

\bibitem[{{Van Allen} {et~al.}(1974){Van Allen}, {Baker}, {Randall}, \&
  {Sentman}}]{vanallen74}
{Van Allen}, J.~A., {Baker}, D.~N., {Randall}, B.~A., \& {Sentman}, D.~D. 1974,
  J. Geophys. Res., 79, 3559

\bibitem[{Webber {et~al.}(2007)Webber, Higbie, \& McCracken}]{webber07}
Webber, W., Higbie, P., \& McCracken, K. 2007, J. Geophys. Res., 112, {A10106}

\end{thebibliography}
%\bibliography{C:/DATA__/papers/usoskin_all}

\begin{appendix}

\section{Reconstructed integral spectra}
\label{App:A}
Event-integrated fluences for the analysed GLE events are presented in the plots below.
Each plot is similar to Figure~\ref{Fig:fit} of the main text in both style and notations.
The GLE number and date are shown on the top of the plots.
Symbols represent data from different sources, as specified in the legend:
 blue crosses and vertical lines correspond to data from neutron monitors and upper estimates,
 respectively, along with their 68\% confidence intervals \citep{usoskin_GLE_20};
 coloured dots correspond to {space-borne}/ionospheric data with the source indicated in the legend
 (`this work' refers to Table~\ref{T:GOES_data});
 and the dark curve depicts the best-fit modified {Band} function (Equations~\ref{Eq:MBF_1} and \ref{Eq:MBF_2};
 exact values of the parameters are available in Table~\ref{T:params}),
 while the light grey shading represents the 68\% confidence interval for the
 fits (see Step 3 of Section~\ref{Sec:fit}).

\pagebreak
.
\newpage
%Added by TeX Support

\includepdf[pages=-,
 nup = 2x3,
 pagecommand*={\null\vfill\captionof{figure}{Integral fluences of SEP reconstructed for GLEs considered in this work (the GLE number and date are given in the header of each panel). Notations are similar to  Fig.~\ref{Fig:fit} of the main text.}},
 % Set the [continued] caption on second and subsequent pages
 % Use '\captionof*{...} to avoid duplicate entries in listoffigures
 %pagecommand={\thispagestyle{plain}\null\vfill\captionof{figure}{\figurename~\thefigure: This is the SECOND caption (Continued)}}
 pagecommand={\null\vfill\captionof*{figure}{\textbf{Fig. \thefigure}.  Continued}}
 ]  {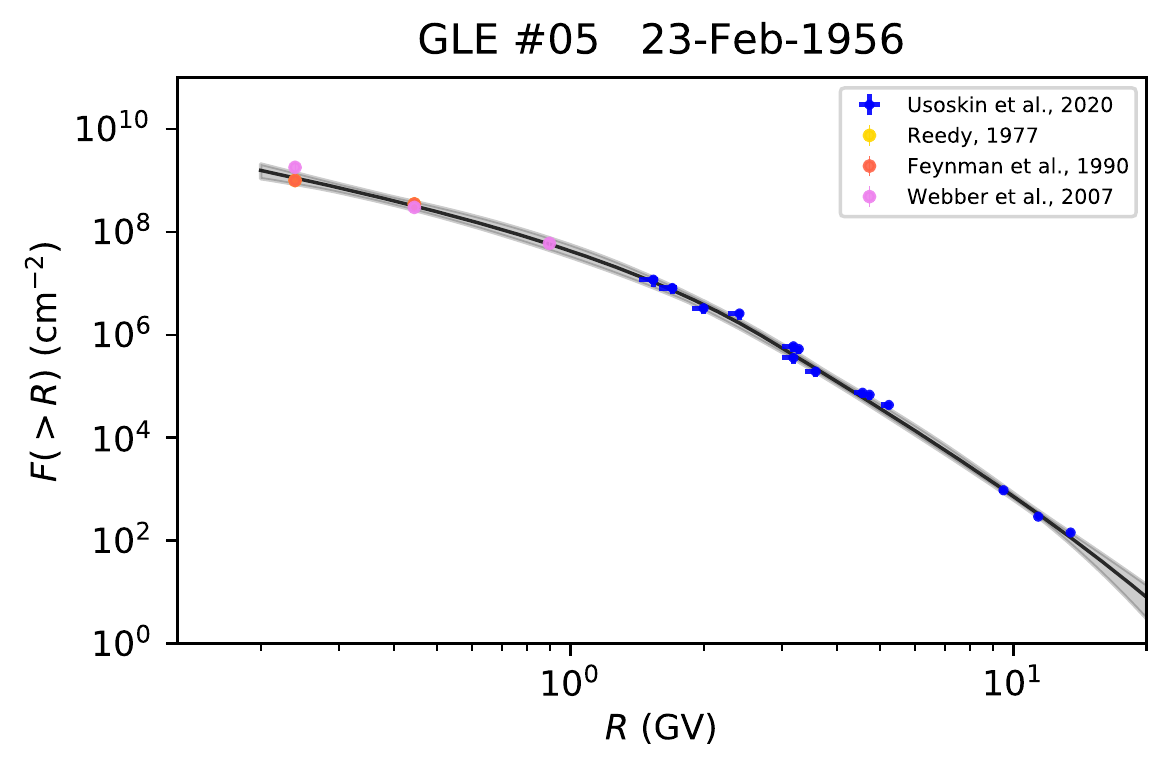}

\end{appendix}

\end{document}